\documentclass[twocolumn,aps,prl,superscriptaddress,floatfix,nofootinbib]{revtex4-2}
\usepackage{amsmath}
\usepackage{amsfonts}
\usepackage{amssymb}
\usepackage{graphicx}
\usepackage{color}
\usepackage{hyperref}
\usepackage{bm}

\newcommand{\beq}{\begin{equation}}
\newcommand{\eneq}{\end{equation}}

\newcommand{\new}[1]{\textcolor{black}{#1}}

\begin{document}

\title{Non-Abelian anyon statistics through AC conductance of a Majorana interferometer }

\author{Andrea Nava}
\affiliation{Institut f\"ur Theoretische Physik,
Heinrich-Heine-Universit\"at, D-40225  D\"usseldorf, Germany}

\author{Reinhold Egger}
\affiliation{Institut f\"ur Theoretische Physik,
Heinrich-Heine-Universit\"at, D-40225  D\"usseldorf, Germany}

\author{Fabian Hassler}
\affiliation{Institute for Quantum Information, RWTH Aachen University, 52056 Aachen, Germany}

\author{Domenico Giuliano}
\affiliation{Dipartimento di Fisica, and INFN, Gruppo Collegato di Cosenza, Universit{\`a} della Calabria, Arcavacata di Rende, I-87036 Cosenza, Italy}
\date{\today}

\begin{abstract}
Demonstrating the non-Abelian Ising anyon statistics of Majorana zero modes in a physical platform still represents a major open challenge in physics. 
We here show that the linear low-frequency charge conductance of a Majorana interferometer containing a floating superconducting island can reveal the topological spin of quantum edge vortices.  The latter are associated with chiral Majorana fermion edge modes and represent ``flying'' Ising anyons. 
We describe possible device implementations and outline how to detect non-Abelian anyon braiding 
through AC conductance measurements. 
\end{abstract}
\maketitle

\emph{Introduction.---}Spectacular experimental progress has recently revealed the fractional exchange statistics of Abelian anyons in the fractional quantum Hall (FQH) regime at filling factor $\nu=1/3$ \cite{Bartolomei2020,Nakamura2020}.  A major open goal is to demonstrate 
the \emph{non-Abelian} anyon braiding statistics expected in more complex
topological phases. Once established, non-Abelian anyons could form the basis of
topological quantum information processing \cite{Nayak2008}.
The simplest non-Abelian quasiparticles are Ising anyons. \new{They are closely related \cite{Barkeshli2019} to Majorana zero modes (MZMs), i.e., non-Abelian topological defects,}
which may be realizable in $p$-wave superconductors (SCs).  The search
for spatially localized MZMs has attracted a lot of recent experimental interest, see \cite{Beenakker2020,Q2023} and references therein. Unfortunately, demonstrations of MZM braiding are still lacking since
disorder-induced conventional fermionic Andreev bound states can mimic many
MZM signatures \cite{Prada2019}.
We note that quantum simulations have reported MZM braiding in digital quantum circuits \cite{Stenger2021,Harle2023}. In the absence of robust physical hardware realizations of MZMs, however, no quantum computational advantage is likely to emerge.  
In the FQH case, the one-dimensional (1D) gapless edge states are chiral (i.e., uni-directional), where
the existence of a bulk gap prevents scattering between different edges.  
As a consequence, anyon braiding is robust against disorder 
\cite{Bartolomei2020,Nakamura2020,Nayak2008,Rosenow2016}.
Similarly, ``flying'' Ising anyons, realized as edge vortices of 1D chiral Majorana fermion edge modes, are expected to be resilient against disorder.  
Edge vortices are composite objects built of a SC vortex (which is one-half of a fermionic flux quantum) and a fermionic excitation \cite{Beenakker2020}.

\begin{figure}[tb]
  \centering
\includegraphics[width = \linewidth]{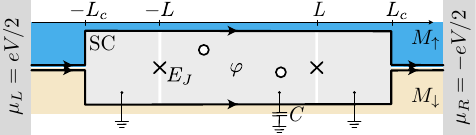}
\caption{Schematic setup. A TI surface is partially covered by oppositely magnetized ferromagnets (M$_\uparrow$ and M$_\downarrow$) and superconductors (SCs), such that co-propagating chiral Majorana edge modes (black arrows) flow around the SC region of total length $2L_c$ and width $2W$. 
The incoming charged 1D chiral Dirac channel is converted with unit efficiency into a pair of charge-neutral chiral Majorana edge modes \cite{Fu2009,Akhmerov2009}. This conversion process is protected by the existence of a bulk gap and by the chirality of the modes. Similarly, Majorana modes are converted back to an outgoing Dirac channel. Charge neutrality is ensured by the SC condensate. Josephson line junctions  (indicated by crosses) define a central floating SC island of length $2L$ with dynamical phase $\varphi$, charging energy $E_C=e^2/2C$, and Josephson energy $E_J\gg E_C$. The outer SC parts are  grounded and held at phase $\varphi_L=\varphi_R=0$.
Potentially present bulk vortices harboring localized MZMs are
indicated by circles.  Note that a left-moving Dirac channel (not shown) 
is located at the outer boundary of the magnetically gapped region. The device is connected
to normal electrodes held at chemical potentials $\mu_L=eV/2$ and $\mu_R=-eV/2$, respectively,
where $V$ refers to an AC voltage with frequency $\omega$.
}
\label{fig1}
\end{figure}

How could one observe the non-Abelian anyon statistics of edge vortices in the simplest manner?
To that end, let us first recall that a pair of 1D chiral Majorana modes can be combined to a 1D chiral Dirac fermion
mode. A conceptually simple Majorana interferometer is  possible by proximitizing a topological insulator (TI) surface \cite{Hasan2010} with magnets and a SC, see Fig.~\ref{fig1} (without the central island of length $2L$) \cite{Fu2009,Akhmerov2009,Nilsson2010,Clarke2010,Hou2011,Simon2018,Shapiro2021}. This setup allows for the electrical detection of chiral Majorana edge modes. 
Even though this interferometer has not yet been demonstrated, recent experimental progress shows that the edge states of quantum anomalous Hall insulators \emph{can} be proximitized by SCs \cite{Uday2023,Amet2016,Lee2017}. This advance makes it likely that Majorana interferometers can soon be realized. We here propose to  
 include a central floating island of length $2L$, see the device layout in Fig.~\ref{fig1}, where
 edge vortices will be dynamically created (or annihilated) due to the finite charging energy of the central island at the rate $\Gamma$ in Eq.~\eqref{Gammadef} below. 
In contrast to FQH interferometers \cite{Bonderson2006,Bonderson2007,Fendley2007,Nayak2008,Fendley2009}, 
the two Majorana fermion edge modes (and the respective dynamically generated edge vortices) move in the \emph{same} direction, with speed $v$.  As we show below, this crucial difference to the FQH case 
 opens up novel avenues for probing the non-Abelian statistics of MZMs 
 through charge conductance measurements.  \new{For a non-specialist summary of our results,
 see Sec.~I of the Supplementary Material (SM) \cite{supp}.}

Recent theoretical work \cite{Beenakker2019a,Beenakker2019b,Adagideli2020,Hassler2020,Flor2023} has studied the
 injection of \emph{deterministic} edge vortices through fine-tuned flux pulses,
where signatures of braiding may be detectable through time-domain charge measurements.
However, the dynamics of \emph{quantum} edge vortices ($\sigma$) is richer as their
non-Abelian statistics is mirrored in non-trivial correlation functions \cite{Nilsson2010,Clarke2010,Grosfeld2011,Hou2011,Ariad2017}. 
In particular, the equal-time correlator of two edge vortices has the long-distance form 
\begin{equation}\label{correl}
\langle \sigma(x) \sigma(0) \rangle \propto e^{2\pi i s_\sigma} |x|^{-2h_\sigma}, 
\end{equation}
with the topological spin $e^{2\pi i s_\sigma}= e^{i\pi /8}$ and the conformal dimension 
$h_\sigma = s_\sigma= 1/16$ \cite{Bonderson2007,Fendley2007,Ariad2017}. 
The topological spin $s_\sigma$ is related to moving an anyon ``around itself,'' and measuring its value
thus directly reveals the non-Abelian braiding statistics, see \cite{Kitaev2006} and App.~A in \cite{Bonderson2007}.
Our key prediction is that a measurement of the linear AC conductance $G(\omega)$ 
for the setup in Fig.~\ref{fig1} allows one to read off the non-trivial topological spin of Ising anyons.
Related ideas have been proposed for measuring the exchange statistics of Abelian FQH quasiparticles 
\cite{Schiller2023}.  However, existing theoretical proposals to detect the braiding statistics in non-Abelian FQH phases require shot noise and/or collision experiments \cite{Lee2022},
which are arguably more challenging than measuring an averaged quantity as suggested here.  For the device in Fig.~\ref{fig1},  $G(\omega)$ 
should be evaluated at $\omega\sim \omega_0=\frac{v}{2(L+W)}$, probing the 
interference of edge vortices around the central SC island. 
Such frequencies are expected in the GHz regime. (We put $\hbar=k_B=1$.)

\emph{Setup and key assumptions.---}We study the device in Fig.~\ref{fig1}, which builds on the well-known
proposal of \cite{Fu2009,Akhmerov2009} but includes a floating central SC island of length $2L$. 
The grounded SCs in Fig.~\ref{fig1} \new{are essential since charge conservation then ensures} 
that the Dirac-Majorana conversion processes \cite{Fu2009,Akhmerov2009} remain well defined. For $E_J\gg E_C$,
rare and fast quantum phase slips, $\varphi\to \varphi\pm 2\pi$,
simultaneously affect both Josephson line junctions defining the central floating island. 
The composite edge vortex creation (or annihilation)
rate $\Gamma$ is estimated from an effective circuit description \cite{Schoen1990},
\begin{equation}\label{Gammadef}
    \Gamma \simeq \sqrt{\frac{8}{\pi}} \omega_p \left( \frac{E_J}{2E_C}\right)^{1/4} e^{-\sqrt{8 E_J/E_C}} ,\quad 
    \omega_p= \sqrt{8E_J E_C},
\end{equation}
where the plasma frequency $\omega_p\gg \Gamma$ sets the inverse time scale on which a phase slip happens.
The rate $\Gamma$ defines an effective charging energy which is reduced from the bare value $E_C$ due to the shunting of the Josephson junction \cite{Hassler2011}.  Later on, we also include the effects of an off-set charge 
$n_g$ (in units of $2e$) on the central SC island which can be tuned by a backgate voltage. In the following, we assume $\omega_p\gg \Delta$ with the induced SC pairing gap $\Delta$.  
Phase slips are then basically time-local events which are not affected by the fermionic sector.
Throughout we assume that the strip width satisfies $2W\gg \xi_0=v/\Delta$, 
i.e., upper and lower Majorana edges do not hybridize except at the Josephson junctions. For clarity, we also assume $2L\gg \xi_0$. 
In principle, bulk quasiparticles could also be excited by instantons 
if $\omega_p\gg \Delta$ \cite{Flor2023}, which at low temperatures may cause weak 
decoherence of the edge vortex dynamics \cite{Grosfeld2011}. 
However, we expect that the charging energy suppresses such effects.
We \new{then} neglect above-gap continuum quasiparticles, assuming that all
relevant energies (in particular, $\omega$ and temperature $T$) stay below $\Delta$.
In that case, transport through the interferometer can only proceed via
the Majorana edge modes because of the SC bulk gap.   While we here discuss the case of equal
path length along the upper and lower branches, we comment on the impact of path differences later on.

In order to derive the AC conductance within linear response theory, 
we employ the Euclidean functional integration framework \cite{Altland2010}, eventually followed by an analytic continuation to real frequency $\omega$.  We here sketch the 
key steps  and present analytical results for small vs large $\Gamma$ compared to the scale $v/L$, respectively.
For details, see the SM \cite{supp}.
The case of arbitrary $\Gamma$ could be investigated in future work, e.g., 
by performing quantum Monte Carlo simulations \cite{Moon1993,Leung1995,Buccheri2017},
but a clear physical picture already emerges from the present study. 

\emph{Chiral bosonization.---}We first proceed along standard steps and 
combine both Majorana edge modes to a single chiral Dirac fermion mode.
This Dirac channel is then bosonized \cite{Delft1998,Altland2010} using a chiral boson field,
$\phi(x,\tau)$, with the 1D coordinate $x$ running along the edge and the imaginary time $0\le \tau\le \beta=1/T$.  The Dirac-Majorana conversion points are then located at $x=\pm (L_c+W)$, and 
the Josephson line junctions are at $x=\pm L$.
For now, we assume that there are no bulk vortices, but we include their effects later on.
It is convenient to define the boson field combinations 
\begin{eqnarray}\label{wpm}
    w_+ (\tau) & = &\frac{\phi(L_c + W,\tau)+\phi(-L_c-W,\tau)}{2},\\
    \nonumber
    w_- (\tau) & = & \frac{\phi(L,\tau)-\phi(-L,\tau)}{2}.
\end{eqnarray} 
The electrical current operator can then be computed from 
the imaginary-time expression \cite{supp}
\begin{equation}\label{cur0}
    I(\tau) = - \frac{ie}{\pi} \frac{dw_+(\tau)}{d\tau}.
\end{equation}
The Euclidean action, $S=S_0+S_\lambda+S_\Gamma+S_V$,  contains four pieces. 
First, the ``free'' action is given by \cite{Fendley2009}
\begin{equation}
    S_0=\frac{1}{4\pi}\int_0^\beta d\tau \int_{-\infty}^\infty dx \, \partial_x\phi(x,\tau)
    \, [i\partial_\tau+v\partial_x]\phi(x,\tau).
\end{equation}
Second, inter-edge fermion tunneling with (real-valued) strength $\lambda_{1,2}$ 
at the junctions at $x_1=-L$ and $x_2=L$ is described by 
\begin{equation}\label{slambda}
    S_\lambda = \sum_{j=1,2} \frac{v\lambda_j}{2\pi} \int d\tau \,  \partial_x \phi(x_j,\tau).
\end{equation}
This term is exactly marginal under renormalization group (RG) transformations and can be absorbed by 
a unitary transformation \cite{Delft1998}.  
Third, edge vortex tunneling at $x=x_{1}$ and $x=x_2$ represents an RG-relevant perturbation. 
In bosonized language, such processes can be described by the operator 
$\sigma_1(x)\sigma_2(x)  = S^- e^{i\phi(x)/2} + {\rm H.c.},$
where the spin-$1/2$ operators $S^\pm=S_x\pm iS_y$ ensure the proper fusion channel  \cite{Fendley2007}.
The Ising anyon fusion rule, $\sigma \times \sigma\sim I+\psi$ \cite{Kitaev2006}, implies that one can either end up 
in the vacuum ($I$) or create a neutral fermion $(\psi)$.   
However, the latter case requires an additional degree of freedom in the junction to accommodate the 
fermion parity change.  
We here consider featureless junctions, where the ``spin'' operator merely represents a bookkeeping prescription \cite{Fendley2007}.  
The action $S_\Gamma$ then describes the composite creation of edge vortices at $x_1$ and $x_2$,
\begin{equation}\label{sgamma}
    S_\Gamma =\Gamma  \int_0^\beta d\tau\,\cos\left[w_-(\tau) +4\pi S_z s_\sigma + 2\pi n_g \right],
\end{equation}
where $n_g$ is the off-set charge parameter on the central SC island. The conserved ``spin'' value $S_z=\pm 1/2$ labels the total fermion parity sector \cite{Fendley2007} corresponding to the fusion outcome $\pm e/2$ when the edge modes are combined down-stream \cite{Adagideli2020,supp}. 
The quantity $s_\sigma=1/16$ governing the topological spin enters via the chiral boson commutator 
algebra \cite{Delft1998}. Note that the topological spin here contributes an effective charge which shifts $n_g$ 
by $\pm s_\sigma$ \cite{Ariad2017}.  Finally, to include the voltage sources in Fig.~\ref{fig1}, 
we employ $S_V = -\frac{e}{2\pi} \int d\tau \, V(\tau)  w_+(\tau)$,
where $V(\tau)$ is used for extracting the AC conductance \cite{Furusaki1993,Egger1998,Giuliano2022,Buccheri2022}.

\emph{Linear AC conductance.---}In the linear response regime, the AC 
conductance $G(\omega)$ then follows by analytic continuation, $-i\Omega\to \omega+i0^+$ 
with Matsubara frequencies $\Omega>0$ \cite{Altland2010}, from the function 
\begin{equation}\label{kubo}
    {\cal G}(i\Omega) = (-1)^{n_v} \frac{e^2 }{2\pi^2 }\, i\Omega \, \langle \tilde w_+(-i\Omega) \tilde w_+(i\Omega) \rangle,
\end{equation}
where the average is taken with the above action for $S_V\to 0$, 
and $\tilde w_+(i\Omega)=\int_0^\beta d\tau e^{-i\Omega \tau} w_+(\tau)$. In Eq.~\eqref{kubo} we also 
took into account the effects of $n_v$ static bulk vortices located far away from the edges.
Following standard arguments, they cause a $(-1)^{n_v}$ prefactor in the conductance \cite{Fu2009,Akhmerov2009,Simon2018}.
At this stage, one can integrate out all bosonic degrees of freedom apart from $w_\pm (\tau)$, 
reminiscent of the resonant tunneling problem in a Luttinger liquid \cite{Kane1992,Furusaki1993}.
The only nonlinearity in the action then comes from $S_\Gamma$ in Eq.~\eqref{sgamma}, which is an RG-relevant perturbation for  $\Gamma<{\rm max}(T,v/L)$.

For $\Gamma=0$, we recover the result of \cite{Fu2009,Akhmerov2009}, $G(\omega\to 0)= (-1)^{n_v} \frac{e^2}{2\pi}$. 
By explicit calculations \cite{supp}, we find that neither edge vortex tunneling 
(for arbitrary values of $\Gamma$) nor fermion tunneling ($\lambda_{1,2}$) are able to
change the DC conductance from the above value for $G(0)$. 
The physical reason for this remarkable effect can be traced to the chiral anomaly \cite{Altland2010} of the Majorana  edge modes. In our context, the anomaly implies that during 
the time $\tau=2\pi/eV$, exactly one fermion will be ``generated'' in the edge modes. 
This fixes the DC electric current to  $I = (-1)^{n_v} e/\tau =G(0)V$.  
Next, for finite but low frequency $\omega \lesssim \omega_0$, we separately 
study $G(\omega)$ for small and large $\Gamma$, respectively, where analytical progress is possible.
(We recall that the SC strip width $2W$ enters the frequency scale $\omega_0$.)

\emph{Weak coupling regime.---}For small $\Gamma\ll v/L$, we perform perturbation theory in $\Gamma$. 
To first order in $\Gamma$, we obtain the low-frequency conductance as
\begin{equation}\label{lowfreqcond}
G(\omega) = G(0) + i (-1)^{n_v} (L_{\rm kin}-C_{\rm eff}) \omega  + {\cal O}(\omega^2),
\end{equation} 
with the kinetic inductance $L_{\rm kin}=\frac{e^2}{\pi v}(L_c+W)$ of the chiral Majorana edge modes.
The leading contribution due to edge vortices appears through an \emph{effective capacitance},
\begin{equation}\label{eq:ceff}
    C_\text{eff} = \Gamma  \frac{e^2L^2 }{2v^2} \cos (\alpha - 4 \pi S_z s_\sigma)
    \left[ \frac{D_0}{T} \sinh ( 2\pi T L/v)  \right]^{-4 h_\sigma}, 
\end{equation}
where $D_0$ is a high-energy bandwidth (corresponding to the bulk gap) and 
$\alpha =  \frac{\pi}{4} (\lambda_1 + \lambda_2)-2\pi n_g$ is a phase shift due to fermion tunneling processes, see Eq.~\eqref{slambda}, and due the off-set charge $n_g$ in Eq.~\eqref{sgamma}.
Second-order terms cause only a small renormalization \cite{supp}  
for $T\agt T^*= {\rm max}\left\{ v/L, \Gamma (\Gamma/D_0)^{1/3}\right\}$, but
perturbation theory breaks down at temperatures below $T^*$.
The effective capacitance $C_\text{eff}$ acts in parallel to the standard conduction
channel due to chiral Majorana edge modes.   As a result, measurements of 
the phase delay $\delta_{\rm ph}=\tan^{-1}\left[\frac{2\pi \omega}{e^2} \left(C_\text{eff}-L_{\rm kin}\right)\right]$ between current and voltage can give access to $C_\text{eff}$; for details, see the SM \cite{supp}.

\begin{figure}[tb]
  \centering
  \includegraphics[width = \linewidth]{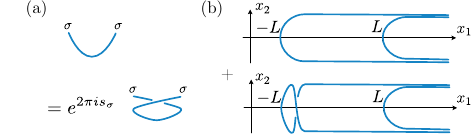}
  \caption{(a) Schematic illustration of how in the algebraic theory of anyons, 
  see App.~E.3 in \cite{Kitaev2006}, 
  the topological spin $e^{2\pi i s_\sigma}$ relates the direct pair creation of edge vortices in a Josephson line junction to the indirect process with interchanged edge vortices. 
  (b) The setup in Fig.~\ref{fig1} (schematically shown in the $x_1$--$x_2$ plane)
  allows one to detect the topological spin since an equal-weight superposition of direct and 
  indirect processes is produced. 
  These processes correspond to a Josephson vortex encircling the central island in the (anti-)clockwise direction, respectively.
  The interference of these processes depends on the topological spin through the relation shown in (a).
  }\label{fig2}
\end{figure}

\begin{figure}[tb]
  \centering
  \includegraphics[width = \linewidth]{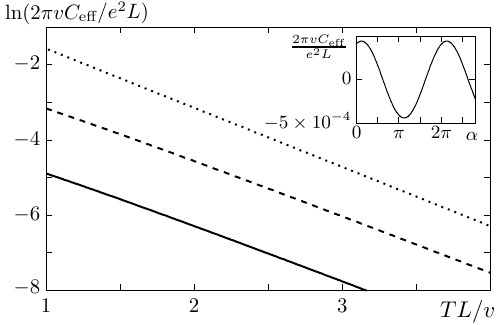}
  \caption{Logarithm of the effective capacitance $C_{\rm eff}$ in Eq.~\eqref{eq:ceff} vs temperature (in units of $v/L$) in the small-$\Gamma$ limit.
  Results are shown for $\Gamma L/v = 0.01$ (solid line) and $\Gamma L/v = 0.1$ (dashed line), 
  with $\alpha=0.2\pi$, $\Gamma/D_0=10^{-3}$, and $S_z=1/2$. 
  From the slope of these curves, which can be inferred through measuring the temperature dependence of $C_{\rm eff}$, the conformal dimension $h_\sigma=1/16$ can be extracted. 
  For comparison, the curve $\ln(2\pi vC_\text{eff}/e^2L) = -8\pi h_\sigma T L/v $ (dotted line) is also indicated. In the inset, we highlight the dependence of $C_{\rm eff}$ 
  on the phase $\alpha$, e.g., when changing $n_g$, assuming $\Gamma L/v = 0.01$, $T L/v=3$, and otherwise the same parameters. 
  The oscillations, in particular the offset, originate from the topological spin $e^{2\pi i s_\sigma}$ of the edge vortices and  are 
  connected to non-Abelian braiding, see also Fig.~\ref{fig2}.}
  \label{fig3}
\end{figure}

From Eq.~\eqref{eq:ceff}, the effective capacitance comes with the scaling dimension $4h_\sigma$ since  $S_\Gamma$ in Eq.~\eqref{sgamma} corresponds to the simultaneous creation of four edge vortices.
These are generated at each intersection of a Josephson 
line junction with a Majorana edge mode in Fig.~\ref{fig1}.
The non-Abelian statistics of the edge vortices appears at several points in Eq.~\eqref{eq:ceff}. 
In particular, $C_{\rm eff}$ depends on the topological spin $e^{2\pi i s_\sigma}$ 
through the $\cos(\alpha - 4\pi S_z s_\sigma)$ factor.
An observation of the oscillatory dependence on $\alpha$ 
could provide direct evidence for non-Abelian anyon braiding, 
as illustrated in Fig.~\ref{fig2} and in the inset of Fig.~\ref{fig3}.  
To estimate the amplitude of the capacitance oscillations, we
compute $\Delta C \sim \Gamma  \frac{e^2L^2 }{v^2} \bigl[ \frac{D_0}{T} \sinh ( 2\pi T L/v) \bigr]^{-1/4}$ from Eq.~\eqref{eq:ceff} for typical parameters \cite{footno}. 
Putting $\Gamma\sim 10$\,GHz, $L\sim 1\,\mu$m, $v\sim 10^4$\,m$/$s, $D_0\sim 1$\,K (which is set by the bulk gap), and $T\sim 0.1$\,K, we obtain $\Delta C\sim 2$\,fF as rough order-of-magnitude estimate.  
Such capacitance changes can be detected by embedding the sample in an LC circuit where capacitance changes of the order of 0.1\,fF have been reported \cite{Droscher2010}.
One may tune $\alpha$ by changing the off-set charge $n_g$ via a backgate voltage \cite{Altland2010}.
By comparing the capacitance value $C^0_\text{eff}$ for $\alpha=0$ 
to the maximum value $C_\text{eff}^*$ (e.g., for $\alpha = 4\pi S_z s_\sigma$), 
the (absolute value of the) topological spin follows from  
$\cos(2\pi s_\sigma)= C^0_\text{eff}/C^*_\text{eff}$.  
Within the validity range of perturbation theory, the capacitance \eqref{eq:ceff} scales as $C_\text{eff} \propto L^2 T^{4 h_\sigma} e^{-8\pi h_\sigma T L/v }$.  As shown in the main panel of Fig.~\ref{fig3},
the conformal dimension $h_\sigma$ of the edge vortices can be measured through the
temperature dependence of $C_\text{eff}$.

\begin{figure}[tb]
  \centering
  \includegraphics[width = \linewidth]{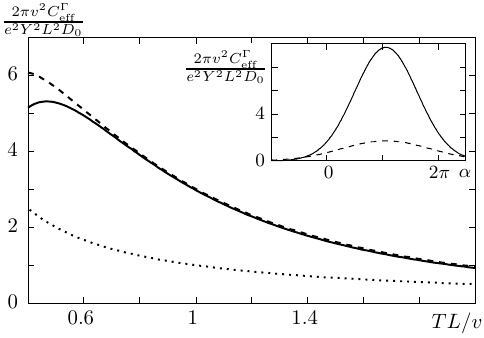}
  \caption{Effective capacitance $C_{\rm eff}^\Gamma$ in Eq.~\eqref{eq:ceffg} vs temperature in the large-$\Gamma$ limit. Results are shown for $v/(LD_0) = 0.1$ (solid line) and $v/(L D_0) =0.01$ (dashed line), with $\alpha = 0.2\pi$ and $S_z=1/2$. The approximate result $C_\text{eff}^\Gamma \simeq  
  \frac{e^2}{2\pi} \frac{Y^2L D_0}{v T}$  (dotted line) is approached for
  $T\gg v/L$. The inset shows $C_{\rm eff}^\Gamma$ vs $\alpha$ for $v/(L D_0) = 0.01$, which is suppressed by increasing   temperature from  $T= v/L$ (solid line) to $T=2v/L$ (dashed line). 
  }\label{fig4}
\end{figure}

\emph{Conductance at strong coupling.---}Next we turn to the regime $\Gamma\gg v/L$, 
where we resort to an instanton calculus \cite{Kane1992,Furusaki1993,Altland2010}. 
For $\Gamma\to \infty$, the field $w_-(\tau)$ is pinned to 
$w_{-,m}  =  \alpha  - 4\pi S_z s_\sigma + \pi (2m+1)$ with integer $m$.
At finite but large $\Gamma$, the leading conductance contributions arise from
(anti-)instanton trajectories interpolating between $w_{-,m}$ with $m\to m+1 \, (m\to m-1)$, respectively. 
These are essentially pointlike objects and form a dilute gas with
 fugacity $Y\propto e^{-\Gamma/D_0}$.  Since instantons are RG-irrelevant, significant
 corrections to the conductance arise only for $T\agt v/L$.
We then obtain $G(\omega)$ as in Eq.~\eqref{lowfreqcond} but with $L_{\rm kin}\to L_{\rm kin}^\Gamma$ and
$C_{\rm eff}\to C_{\rm eff}^\Gamma$ \cite{supp}.  We here quote only the capacitance,
\begin{equation}\label{eq:ceffg}
    C_\text{eff}^\Gamma \!\simeq\! \frac{e^2 Y^2}{2\pi}\!\sqrt{\frac{v T}{L }}
    \frac{T^4}{D_0^6} \int\!dz\frac{z^2  
    e^{-(v/2\pi T L) (2z + \pi + \alpha -2  \pi  S_z s_\sigma )^2}}{\left|\sin^8(z +i T/D_0)\right|}.
\end{equation} 
To leading order in $D_0^{-1}$, for $T\gg v/L$, we obtain  $C_\text{eff}^\Gamma \simeq  \frac{e^2}{2\pi} \frac{Y^2L D_0}{v T}$. These results suggest that the strong-coupling limit is less favorable for
detecting non-Abelian statistics.  In fact, Fig.~\ref{fig4} shows that the temperature dependence
of $C_{\rm eff}^\Gamma$ does not give direct access 
to $h_\sigma$ anymore.  Moreover, a
topological spin contribution appears only at subleading order in the small parameter $v/(LT)$, 
see the inset in Fig.~\ref{fig4}.

\emph{Discussion.---}The device in Fig.~\ref{fig1} can reveal the elusive Ising anyon braiding statistics through measurements of the linear AC conductance.  For small edge vortex production rate $\Gamma$ 
in Eq.~\eqref{Gammadef}, which is a natural regime for experimental realizations, one expects optimal working conditions.Our theory assumes equal path length for both arms of the setup in Fig.~\ref{fig1}.
Using the results of \cite{Fu2009,Akhmerov2009,Nilsson2010,Clarke2010,Hou2011,Simon2018,Shapiro2021,Wei2023},
we estimate the path length difference $\Delta L$, above which one can expect qualitative changes to our results, from the ``size'' of an edge vortex. In time units, the latter is determined by the injection time $t_\text{inj} = \xi_0/(W \dot\phi)$ \cite{Beenakker2019b}, where $\dot \phi\approx \omega_p$ with the plasma frequency $\omega_p$ in Eq.~\eqref{Gammadef}. 
We thus obtain $\Delta L\approx  v  \xi_0/(W\omega_p)$. For a practical estimate, let us take 
\new{$W\sim 2\,\mu$m, $\omega_p\sim 50$\,GHz,} $v\sim 10^4$\,m$/$s, and $\xi_0\sim 1\,\mu$m, resulting in \new{$\Delta L\sim 0.1~\mu$m.}
We note that a small velocity mismatch between the Majorana edge modes has a similar effect as $\Delta L$.
Finally, to formulate the theory for arbitrary arm length difference or for more complex devices, methodological advances are needed. It would also be interesting to study the nonlinear AC conductance for our setup, see also \cite{Nava2021a,Nava2023b}.

\begin{acknowledgments}
We thank A. Akhmerov, Y. Ando, C. Beenakker, E. Bocquillon, and I. M. Fl{\'o}r for discussions.
We acknowledge funding by the Deutsche Forschungsgemeinschaft (DFG, German Research Foundation),
Projektnummer 277101999 -- TRR 183 (project C01), under project No.~EG 96/13-1,
and under Germany's Excellence Strategy -- Cluster of Excellence Matter and Light for 
Quantum Computing (ML4Q) EXC 2004/1 -- 390534769.  The data underlying the figures in this work can be 
found at the zenodo site: \url{https://doi.org/10.5281/zenodo.13269567}
\end{acknowledgments}

{\color{black}
\section{Supplemental Material}
\label{SM}

We here provide details on the derivations of our results presented in the main text.
\new{First, in Sec.~I, we provide a basic summary of the considered setup and briefly 
outline the main ideas of our protocol.}  
In Sec.~II, we discuss the imaginary time approach to the linear AC conductance of the 
Majorana interferometer and use it to derive the low-frequency conductance in the limits of 
small and large edge vortex production rate $\Gamma$, respectively. In Sec.~III, 
we provide arguments as to why the DC conductance of the interferometer is not affected
by arbitrary values of $\Gamma$, and thus given by its well-known $\Gamma=0$ value.
Equation (X) in the main text is referred to as Eq.~(MX) below.

\new{
\section{I. Experimental setup and basic ideas}\label{intro}
}

\begin{figure}[b]
\includegraphics[width=\columnwidth]{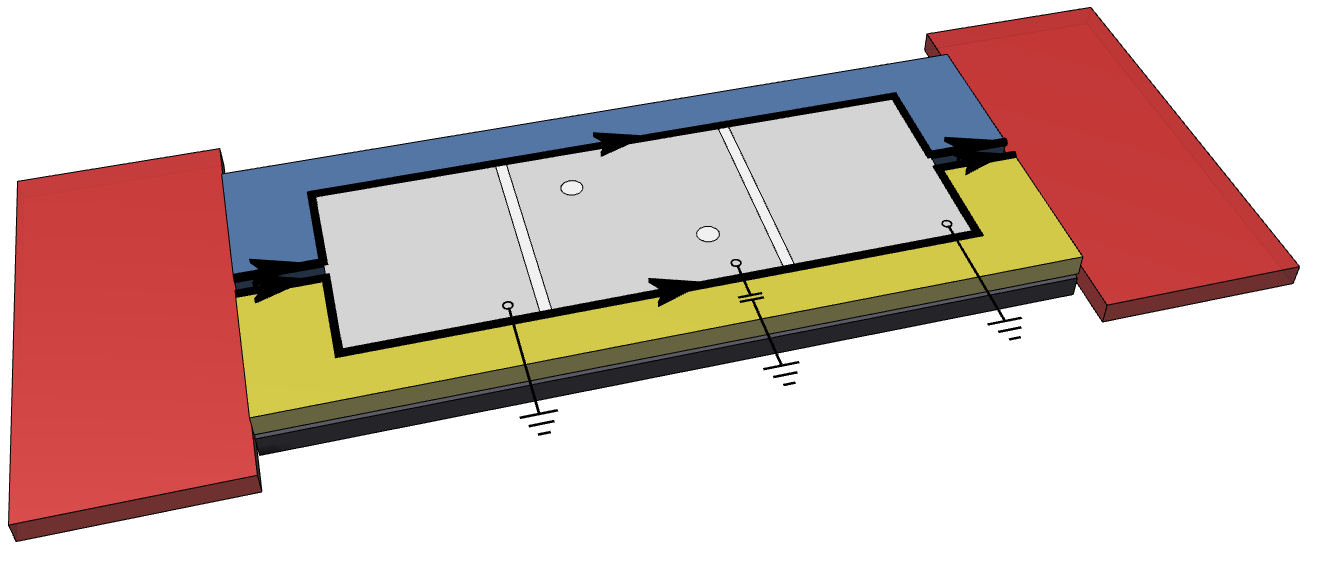}
\caption{\new{Schematic three-dimensional (3D) view of the considered experimental device for observing non-Abelian braiding of Ising anyons via the AC conductance.  The red blocks represent left and right metallic leads which are employed to measure the conductance through the Majorana interferometer.  The device is composed of a 3D topological insulator, e.g., Bi$_2$Se$_3$, which harbors a single two-dimensional (2D) gapless Dirac fermion cone at its surface.  
By depositing either conventional $s$-wave superconductors or magnetic films on the surface, one gaps out the Dirac fermions in different manners \cite{Fu2009,Akhmerov2009}.  At the interface between both gapped regions, a chiral Majorana fermion mode 
exists.  By using two oppositely magnetized films (blue and green regions) surrounding a superconducting region, one can realize an incoming chiral Dirac fermion channel which splits into copropagating chiral Majorana modes eventually merging to an outgoing Dirac channel.  The central superconductor is composed of two outer grounded
pieces and a central floating island defined by a pair of Josephson line junctions.  Vortices, potentially present in the superconducting regions (shown as circles), harbor additional static Majorana zero modes. }}\label{fig:s0}
\end{figure}

\new{Here we give a short summary for non-specialists of the main ideas behind observing non-Abelian braiding of Ising anyons via the AC conductance. A detailed discussion is contained in 
the main text and in the remainder of the SM. 
The schematic experimental device is shown in Fig.~\ref{fig:s0}, see also Fig.~1 in the main text, where one 
employs the 2D massless Dirac fermion surface state of a strong topological insulator such as Bi$_2$Se$_3$. 
This state can be gapped out in different manners by depositing either a conventional superconductor (central region) or magnetic materials.  By using two oppositely magnetized materials, one can create the chiral Majorana interferometer
proposed in Refs.~\cite{Fu2009,Akhmerov2009}, where an incoming chiral Dirac fermion channel splits into a pair of copropagating chiral Majorana fermion modes which finally fuse to an outgoing chiral Dirac channel.  
We go beyond the proposal in Refs.~\cite{Fu2009,Akhmerov2009} by adding a central floating superconducting island, which can be defined by a pair of Josephson line junctions.  The outer grounded superconductors are necessary to make sure that the conversion between chiral Dirac fermions and pairs of chiral Majorana fermion modes is well defined.  The charging energy and the Josephson energy of the central floating island then determine the rate $\Gamma$ for 
the creation or annihilation of phase slips, $\varphi\to \varphi\pm 2\pi$, of the superconducting phase $\varphi$ on the central island.   Such a phase slip simultaneously generates (or annihilates) four edge vortices  at the endpoints of the Josephson line junctions.   Edge vortices \cite{Beenakker2020}
can be thought of as coreless domain walls for a chiral Majorana mode, associated with a sign change at the position of the edge vortex.  In analogy to the Jackiw-Rebbi zero-energy fermion mode bound to a mass kink 
for one-dimensional chiral Dirac fermions, and by recalling that a Majorana fermion is essentially ``half'' a Dirac fermion, edge vortices bind a Majorana zero mode but do not come with additional bound states (as is usually the case for vortices in topological superconductors).      
Edge vortices are therefore representatives of Ising anyons.  In particular, they are characterized by a topological spin
$e^{2\pi i s_\sigma}$ with $s_\sigma=1/16$, which encodes their  non-Abelian braiding statistics \cite{Kitaev2006}.
We note that these Ising anyons are moving along the edge with the edge velocity, in contrast to the commonly considered static variants.  However, the chirality of the edge modes and the absence of additional bound states in a coreless edge vortex imply a much higher robustness against disorder effects compared to their static counterparts.}

\new{To lowest order in $\Gamma$, an effective capacitance term appears in the low-frequency part of the AC conductance.  This effective capacitance can be measured via the phase delay between the voltage and the current, and it is predicted to feature an oscillatory dependence on a parameter $\alpha$. The latter parameter can be tuned experimentally by varying external gate voltages.  We predict that the topological spin manifests itself in a phase shift of the oscillations due to the interference of different contributions, 
see Fig.~2 in the main text.
Since the topological spin is a direct consequence of non-Abelian braiding, a measurement of the AC conductance at low frequencies can probe the braiding statistics of Ising anyons.  Such experiments are expected to be simpler than the shot noise or anyon collision proposals \cite{Bonderson2006,Lee2022} for fractional quantum Hall liquids with non-Abelian statistics.  
}\\

\section{II. Imaginary time approach to the linear AC conductance}
\label{imaginary}

We here present our imaginary time approach for computing the
frequency-dependent linear conductance $G(\omega)$
of the Majorana interferometer shown in Fig.~1 of the main text. We employ this 
technique to derive $G(\omega)$ in the zero-frequency (DC) limit, $G(0)$,  
  as well as the leading contribution in $\omega$ to the AC conductance.  After a summary of the general structure of the effective
action in Sec.~II A, we provide analytical results for
the low-frequency AC conductance for small edge vortex tunneling (EVT) rate  $\Gamma$ in 
Sec.~II B, and subsequently for large $\Gamma$ in Sec.~II C.
In Sec.~III, we show that for arbitrary $\Gamma$,  $G(0)$ is not affected by the EVT rate $\Gamma$.

\subsection{A. Derivation of the effective action}
\label{effac}

We first derive the effective Euclidean action ${\cal S}[w_+,w_-]$ governing the
fields $w_\pm (\tau)$ defined in Eq.~(M3).
To that end, we start from the 1D chiral Dirac fermion field
$\Psi (x) = \frac{1}{\sqrt{2}} [ \psi_1 (x) + i \psi_2 (x )]$, where
$\psi_1 (x)$ and $\psi_2(x)$ are chiral Majorana fermion operators living on the upper and lower edge of the SC part of the device shown in Fig.~1 of the main text, respectively; see also \cite{Fu2009,Akhmerov2009}. 
Here, $x$ is taken as a coordinate running along the edge, i.e., the 
Dirac-Majorana conversion points are located at $x=\pm (L_c+W)$ and the Josephson line
junctions are at $x=\pm L$.  For $|x|>L_c+W$, the field $\Psi(x)$ describes the Dirac channel.

For bosonizing the Dirac field, we introduce a chiral boson field $\phi(x)$ \cite{Delft1998}, such that
$\Psi (x)$ and the charge density operator  $\rho(x)$ are respectively realized as 
$\Psi (x) = : e^{-i\phi (x)}:$ and $\rho (x) = e \Psi^\dagger (x) \Psi (x) =\frac{e}{2\pi} 
\partial_x \phi (x)$. Here the double colons
$:\: :$ denote normal ordering with respect to the ground state of the bosonic theory. 
Within the imaginary time ($0\le \tau\le \beta=1/T$ for temperature $T$) 
framework \cite{Altland2010}, 
the free Euclidean action $S_0$ for $\phi (x,\tau)$ and the term $S_\lambda$ describing edge fermion tunneling at the Josephson line junctions at $x=\pm L$ are then given by Eqs.~(M5) and (M6), respectively.  EVT processes may happen at $x_1=-L$ and $x_2=L$ if phase slips take place. 

Following \cite{Adagideli2020}, we can model the effect of the Josephson junction with a time-dependent phase difference $\varphi(t)$ by a phase field, 
\begin{equation}
    \Lambda(t) \approx \arccos\tanh\left[ (W/\xi_0) \cos(\varphi(t)/2) \right],
\end{equation}
where $W$ is the width of the Josephson line junction and $\xi_0$ the superconducting coherence length. 
Note that for a phase slip, where $\varphi$ advances by $2\pi$, we find that $\Lambda$ increases by $\pi$. In particular, if $W \gg \xi_0$, the increase of $\Lambda$ is close to step-like. The EVT operator at position $x$ is then given by 
\begin{align}
  \mathcal{T}_\sigma(x) &= \sigma_1(x) \sigma_2(x) \nonumber\\
    &= S^- \exp\left( - \frac{i}{e} \int\!dx' \,\tilde\rho(x')  \Lambda(x-x')\right) + \text{H.c.} \nonumber\\
                  &= S^- e^{- 2i\int\!dx' n_s(x')\Lambda(x-x')  - \frac{i}{2\pi} \int\!dx' \phi(x') \partial_x \Lambda(x-x')} \nonumber\\
                        &\quad+ \text{H.c.},\label{eq:vt}
\end{align}
 with the quantum edge vortex operator $\sigma (x)$ in Eq.~(M1).  
Note that, for convenience and different from  \cite{Adagideli2020}, we write this operator here in the Schr\"odinger picture. The charge density operator appearing in Eq.~\eqref{eq:vt} is given by 
$\tilde\rho(x)= \rho(x) + 2 e n_s(x)$, with the density $n_s$ of Cooper pairs. 
The spin degree of freedom, with $S_z = \pm \frac12$ and $S^\pm=S_x\pm i S_y$, 
takes care about the fusion outcome $\pm e/2$ of the charge flowing downstream. In particular, an increase (decrease) of the phase by $2\pi$ leads to the charge $e/2$ ($-e/2$) \cite{Hassler2020}.

Within the bosonization framework of Fendley \emph{et al.}\cite{Fendley2007}, an elementary EVT process at location $x$ is described by the Hermitian coupling operator 
\begin{equation}\label{fendl1}
   {\cal T}_\sigma (x )= \sigma_1 (x) \sigma_2 (x) = S^-e^{\frac{i}{2} \phi (x)} + {\rm H.c.} .
\end{equation}
 As explained in the main text and in \cite{Fendley2007}, we assume that the ``spin'' $S_z=\pm 1/2$ appearing in Eq.~\eqref{fendl1} is conserved. (If parity-changing processes  are present, e.g., due to quasi-particle poisoning or related effects, a finite ``spin'' lifetime may be possible.  However, in practice, we expect that this lifetime will be very long.)
Interestingly, we obtain Eq.~\eqref{fendl1} from the more general expression \eqref{eq:vt} by 
assuming $W\gg \xi_0$ such that $\partial_x\Lambda(x) = \pi \delta(x)$. In the main text, we have used Eq.~\eqref{fendl1}.

 In our case, a phase slip happens at the rate $\Gamma$ and affects  both junctions simultaneously. 
 The corresponding composite Hermitian coupling operator describing EVT is obtained from a symmetrized coupling,
 $H_{\rm \Gamma}=\frac{\Gamma}{2} \{{\cal T}_\sigma (x_1 ),{\cal T}_\sigma (x_2)\}$, 
 where $\{ \cdot,\cdot\}$ denotes the anticommutator.
 This expression implies the action $S_\Gamma$ in Eq.~(M7).
 
 We here assume that the (within the linear response regime) voltage bias $V(t)\propto V_\omega e^{i\omega t}$ is symmetrically 
 applied between the leads at $x\leq -(L_c +W)$ and $x\geq (L_c +W)$, see Fig.~1 of the main text. The voltage  therefore couples to the charge imbalance operator
 \begin{eqnarray}\nonumber
      \Delta Q &=& \int_{-\infty}^{-(L_c+W)} \: dx \: \rho (x) - \int_{L_c+W}^\infty \: dx \: \rho (x) \\
      &=& \frac{e}{2\pi}[\phi(L_c +W)+\phi(-L_c-W)].
 \end{eqnarray} 
 The charge current operator $I$ accordingly follows
 as $I=\frac{d \Delta Q}{dt}$. 
 Apparently, the whole system dynamics is then essentially determined by the 
 two bosonic field combinations $w_\pm(\tau)$ in Eq.~(M3). Motivated by this observation, we resort to an
 effective description in terms of $w_\pm(\tau)$ only. 
 The corresponding imaginary-time expression for $I(\tau)$ is given by Eq.~(M4), 
 and the action contribution due to the voltage is
 $S_V = -\frac{e}{2\pi}  \int d\tau \, V(\tau) w_+(\tau)$,  as specified in the main text.
 
 Within the imaginary time framework, we can derive an effective action ${\cal S}[w_+,w_-]$
 by functional integration over $\phi (x,\tau)$, where the definition of
 $w_\pm (\tau)$ is enforced through bosonic Lagrange multiplier fields.  Integration over the now
 Gaussian field $\phi(x,\tau)$ and, subsequently, over the (also Gaussian) Lagrange multiplier fields
 yields the desired action. As a result, we obtain 
 \beq\label{sw+w-}
 {\cal S} [w_+,w_-] =  {\cal S}_0 [w_+ ,w_- ]
 + S_\Gamma,
 \eneq
 with  $S_\Gamma$ in Eq.~(M7). The ``free'' action is given as a sum over
 bosonic Matsubara frequencies $\Omega$,
 \begin{equation}
 {\cal S}_0 [w_+,w_-]=\frac{T}{2} \sum_{i\Omega} \sum_{a,a'=\pm} \:{\cal K}_{a,a'} (i\Omega) 
 \bar{w}_a(-i\Omega) \bar{w}_{a'}(i\Omega ), 
 \label{appe.1.3}
 \end{equation}
 with the shifted fields
 \begin{equation}
 \bar{w}_a (i\Omega) = \tilde{w}_a(i\Omega) - \frac{ \pi \alpha }{T}\delta_{\Omega ,0}\delta_{a,-}.
 \end{equation} 
We define $\tilde{w}_\pm (i\Omega) = \int_0^\beta d\tau \: e^{-i\Omega \tau} w_\pm (\tau)$ as in the 
main text, and we again use the phase 
\begin{equation}
    \alpha= \frac{\pi}{4}(\lambda_1+\lambda_2) -2\pi n_g,
\end{equation}
arising due to the fermion tunneling action $S_\lambda$ and due to the off-set charge parameter $n_g$
on the central superconducting island.
The kernel in Eq.~\eqref{appe.1.3} has the components 
 \begin{eqnarray}
 {\cal K}_{+,+}(i\Omega) &=& \frac{2|\Omega|}{\pi}  \frac{1-e^{-\frac{2L |\Omega|}{v}}}{\Delta (i\Omega)}  ,
 \nonumber \\
  {\cal K}_{-,-}(i\Omega) &=& \frac{2|\Omega|}{\pi} \frac{1+e^{-\frac{2(L_c+W) |\Omega|}{v}}}{\Delta (i\Omega)}  ,
  \nonumber \\
  {\cal K}_{+,-}(i\Omega )&=&  \frac{2|\Omega|}{\pi} \frac{e^{-\frac{(L_c+W-L)|\Omega|}{v}}-
  e^{-\frac{(L_c+W+L) |\Omega|}{v}}}{\Delta (i\Omega)}   , \nonumber \\
   {\cal K}_{-,+}(i\Omega )&=& - {\cal K}_{+,-}(i\Omega ) , 
   \label{appe.1.4}
   \end{eqnarray}
with the quantity
\begin{equation}
\Delta (i\Omega)= 1-e^{-\frac{2L|\Omega|}{v}}-e^{-\frac{2(L_c+W) |\Omega|}{v}} 
 + e^{-\frac{2(L_c+W-L)|\Omega|}{v}}.
\end{equation}
While our results superficially resemble the theory of resonant tunneling in a Luttinger liquid in \cite{Kane1992,Furusaki1993},
 Eqs.~\eqref{appe.1.3} and \eqref{appe.1.4} differ from the analogous equations in that case in two aspects.
 First, since we apply the bias voltage at $x=\pm (L_c+W)$ while the Josephson line junctions are located at $x=\pm L$, here two different length scales ($L$ and $L_c+W$)  appear in the matrix kernel ${\cal K} (i\Omega)$.
  Second, due to the chiral nature of the boson field $\phi(x,\tau)$, the kernel has nonzero
  off-diagonal elements while it is purely diagonal for the resonant tunneling case \cite{Kane1992,Furusaki1993}.

 \subsection{B. Low-frequency conductance for small $\Gamma$}
 \label{lowf}

We next use the action ${\cal S}[w_+,w_-]$ in Eq.~\eqref{sw+w-} in order to compute,  within linear response theory, the low-frequency 
conductance $G(\omega)$.  This quantity is obtained from the function ${\cal G}(i\Omega)$ in Eq.~(M8)
by analytically continuing to real frequencies, where Eq.~(M8) is evaluated for $V_\omega \to 0$.
For simplicity, in the remainder of the SM, we shall put the factor $(-1)^{n_v}\to 1$ in Eq.~(M8).
For $\Gamma=0$, we obtain
\begin{eqnarray}
G_0 (\omega) &=&-  \frac{\omega e^2}{2\pi}  
 [{\cal K}^{-1}]_{+,+}(-i\Omega \to \omega + i0^+) \nonumber \\
 &=& 
\frac{e^2}{4\pi} \left(1+e^{\frac{2i(L_c+W)\omega}{v}}\right)  . 
\label{appe.2.2}
\end{eqnarray}
\noindent
By expanding to first order in $\omega$, we
obtain the DC conductance $G(0)=e^2/2\pi$ \cite{Fu2009,Akhmerov2009}. In addition, we find a low-frequency contribution proportional to the
kinetic inductance $L_{\rm kin}$ due to the chiral Majorana edge modes, see Eq.~(M9).  

Remarkably, a finite EVT coupling $\Gamma$ will not affect
$G (0)$, as we discuss in the main text and in Sec.~III.
To leading order in $\omega$, however, we find that $\Gamma\ne 0$ implies a finite \emph{effective 
capacitance} ($C_{\rm eff}$) contribution to $G(\omega)$. In order to compute $C_{\rm eff}$, we note that for small $\Gamma$, $G(\omega)$ can be expanded as a perturbation series in $\Gamma$,
\beq
G (\omega)=G_0(\omega) + \Gamma \bar{G}_1 (\omega) + 
\Gamma^2 \bar{G}_2 (\omega) + {\cal O}(\Gamma^3).
\label{appe.2.3}
\eneq 
Defining the inverse kernel functions
\begin{equation} 
g_{a,a'}(\tau)=T\sum_{i\Omega} e^{i\Omega \tau } [{\cal K}^{-1}]_{a,a'}(i\Omega),
\end{equation}
and expanding Eq.~\eqref{appe.2.3} to first order in $\omega$, we obtain
\begin{eqnarray}\nonumber
\bar{G}_1(\omega)&=&-\frac{i e^2L^2\omega}{2 v^2} \cos (\alpha-4\pi S_z s_\sigma) e^{-\frac{1}{2} g_{-,-} (0)} 
\\
&=& -\frac{i e^2L^2\omega}{2 v^2} \cos (\alpha-4\pi S_z s_\sigma) \times \nonumber
\\ &\times& \left[\frac{D_0}{T} \sinh \left( \frac{2\pi TL}{v \beta} \right)\right]^{-4h_\sigma}, \label{appe.2.4} 
\end{eqnarray}
\noindent
where the bulk gap serves as high-energy cutoff $D_0$ and we recall $h_\sigma=s_\sigma=1/16$.  We thus obtain 
$C_{\rm eff}$  in Eq.~(M10). 

The above perturbation theory holds for $T \gtrsim {v}/{L}$.
In addition, from Eq.~(\ref{appe.2.4}), we infer the scaling of the effective ``running'' EVT coupling, 
 $\Gamma_r(T) = \Gamma (T/D_0)^{1/4}$, which
 has to be compared to the typical energy scale associated to thermal fluctuations, $E_{\rm th}\sim T$. To remain 
 consistent with our assumption of small ``bare'' $\Gamma$, we require $\frac{\Gamma_r}{E_{\rm th}} \lesssim 1$.
 We thus arrive at the condition $T\gtrsim T^*= {\rm max} \left\{
v/L , \Gamma\left(\Gamma/D_0\right)^{1/3} \right\}$ for the perturbative regime, as specified in the main text.
 This conclusion is further supported by 
 computing $\bar{G}_2 (\omega)$ in Eq.~\eqref{appe.2.3}. Indeed, performing the calculation as for $\bar{G}_1(\omega)$ and
 retaining again only terms linear in $\omega$, we find
 \begin{eqnarray}
 \bar{G}_2 (\omega)  &=&   -\frac{i e^2L^2\omega}{8 v^2}  \left[\frac{D_0}{T} \sinh \left( \frac{2\pi TL}{v } \right)\right]^{-8h_\sigma} 
   \label{appe.2.5} \\
 &\times&
  \nu_0 \left( \frac{2\pi TL}{v} \right)\cos (2\alpha-8\pi S_z s_\sigma)  ,  \nonumber
    \end{eqnarray}
  \noindent
  where $\nu_0 (z)$ is a non-universal scaling function (its precise form can be written down but is not of interest in our context). Apparently, Eq.~(\ref{appe.2.5}) implies the same scaling of 
  $\Gamma_r$ as in Eq.~(\ref{appe.2.4}), and the corresponding correction to $C_{\rm eff}$ is given by 
  $\Gamma_r^2$ times a scaling function of the dimensionless ratio  $\frac{2\pi TL}{v }$. In addition, we have numerically
  checked that, even for $\Gamma$ as large as $0.1 D_0$, $\bar{G}_2$ satisfies $\bar G_2\ll \bar{G}_1$ throughout the range of relevant values
  of $T$ and $\omega$.  For these reasons, we have neglected second-order contributions in the main text.

Finally, we comment on how to measure $C_{\rm eff}$ through the phase delay between the voltage and the current.
In the main text, we have given the estimate $C_\text{eff} \sim 1\,$fF. We propose to perform the experiment at  $\omega\sim 1\,$GHz $\lesssim \omega_0\sim 10\,$GHz. The phase shift due to the effective capacitance is then given by $\delta_{\rm ph}=\tan^{-1}(2\pi \omega C_\text{eff}/e^2) \sim 5^\circ$. Applying an AC voltage of frequency $\omega$, the resulting AC current through the device leads the voltage by $\approx 5^\circ$, see Fig.~\ref{fig:s1}.

\begin{figure}[tb]
\includegraphics{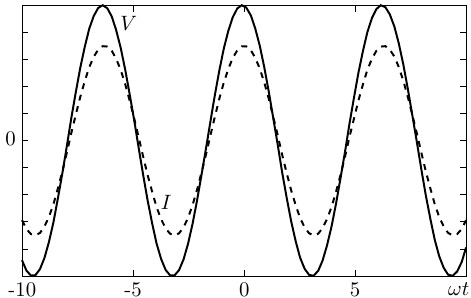}
\caption{The current $I$ (dashed line) leads the  voltage $V$ (solid line), both given in arbitrary units, by approximately $5^\circ$ when operated at frequency $\omega=1\,$GHz due to the edge-vortex tunneling that leads to an effective capacitance.}\label{fig:s1}
\end{figure}

 \subsection{C. Low-frequency conductance for large $\Gamma$ }
 \label{larga}
 
 As a preliminary step towards computing $G(\omega)$ in the large-$\Gamma$ limit, we 
 first show how the pinning of $w_- (\tau)$ to one of the values $w_{-,m}$ with integer $m$ quoted in the main paper, 
 enforced due to the large prefactor $\Gamma$ in the EVT action term $S_\Gamma$ in
  Eq.~(M7), will affect the partition function ${\cal Z}$. 
 From Eq.~(\ref{appe.1.3}) and noting that $S_\Gamma$ only depends on 
 $w_-$, the action remains Gaussian in $w_+(\tau)$, regardless of 
 the value of $\Gamma$. We can therefore integrate over the field $w_+$, thus arriving at an effective 
 action
 \beq
 {\cal S}_- [w_-] = \frac{T}{2} \sum_{i\Omega} {\cal K}_{-,-} (i\Omega) \tilde{w}_-(-i\Omega) \tilde{w}_- (i\Omega)  + S_\Gamma. 
 \label{appe.3.1}
 \eneq
 For $\Gamma\gg v/L$, the contribution to the partition function 
 arising from the $w_-$ sector can be accurately estimated by performing a saddle-point expansion, 
$w_- (\tau) = w_{-,m }+ \delta w_- (\tau)$. 
Neglecting subleading corrections $\propto v /L\Gamma$,
the static (time-independent) saddle-point solutions are given by $w_{-,m}$ in the
main paper,
which minimize $S_\Gamma$.  With $\delta w_-(\tau)$, we can 
then construct instanton--anti-instanton and 
anti-instanton--instanton trajectories, which connect two neighboring static solutions with indices
$m$ and $m\pm 1$, respectively. 

As we find below, instantons are RG-irrelevant perturbations for large $\Gamma$. 
This implies that considering only corrections due to a single instanton--anti-instanton 
pair gives already an accurate approximation for  $G(\omega)$ if $\Gamma\gg v/L$. 
Accordingly, instanton--anti-instanton trajectories connecting ${w}_{-,m}$ to
${w}_{-,m+1}$ and back are realized as
\beq
w_{-,{\rm IA}}^{(m)} (\tau) = {w}_{-,m} + 2 \pi [\theta (\tau - \tau_1 ) - \theta (\tau-\tau_2) ]  ,
\label{appe.3.3}
\eneq
where $\theta(z)$ is the Heaviside step function.
Similarly, anti-instanton--instanton solutions connecting $w_{-,m}$ to 
${w}_{-,m-1}$ and back take  the form 
\beq
w_{-,{\rm AI}}^{(m)}(\tau) = {w}_{-,m} - 2 \pi [ \theta (\tau-\tau_1 ) - \theta (\tau-\tau_2)] . 
\label{appe.3.4}
\eneq
In Eqs.~(\ref{appe.3.3}) and \eqref{appe.3.4}, $\tau_1$ and $\tau_2$ refer to the center-time locations of the (anti-)instanton, taken in the bounds $0<\tau_1<\tau_2 < \beta$.  Each (anti-)instanton is governed by the fugacity $Y\sim e^{-\Gamma/D_0}$, where $D_0^{-1}$ 
determines the width of the (anti-)instanton transition in imaginary time.  
(For simplicity, we have put $D_0^{-1}\to 0$ in  Eqs.~\eqref{appe.3.3} and \eqref{appe.3.4} when using
the Heaviside step function.)

For the resulting contribution to the 
partition function due to $w_-$, we thereby obtain the estimate 
\begin{widetext}
\begin{equation} \label{appe.3.5} 
 {\cal Z}   =   {\cal N} \sum_{m=-\infty}^\infty  \left(
  e^{-\frac{\beta v}{2 \pi L} \left[  2 \pi \left(m + \frac{1}{2} + \frac{\alpha}{2\pi}  - S_z s_\sigma \right)\right]^2} 	+
 Y^2 \beta D_0^2 \int_0^\beta d \tau   e^{-\frac{2 v \beta}{\pi L} \left[\left(\frac{\pi \tau}{\beta} \right)+
\frac{\alpha}{2} +  \pi \left(m+\frac{1}{2} -S_z s_\sigma \right) 
\right]^2-{\cal W} \left(\frac{\pi \tau}{\beta} \right) } \right),  
 \end{equation}
 \end{widetext}
 where the factor ${\cal N}$ takes into account contributions due to fluctuations around the saddle-point
 solution.  (For our discussion below, however, the precise result for ${\cal N}$ is irrelevant.) 
 Moreover, we use the function
\beq
 {\cal W}(z) = 8 \sum_{n=1}^\infty  \frac{1-\cos \left(  2  n z \right)}{n
\left(1 - e^{-\frac{4 L \pi n}{v \beta} } \right)}. 
\label{appe.3.6}
\eneq
We note that in deriving Eq.~(\ref{appe.3.5}), the explicit integration 
over the instanton--anti-instanton center-of-mass coordinate $(\tau_1+\tau_2)/2$
(normalized to the instanton size $D_0^{-1}$)
results in a factor $\beta D_0$ contributing to the term $\propto Y^2$.

In the main text, we focus on the regime $T\agt v/L$, where Eq.~(\ref{appe.3.6}) can be approximated as
\beq
 {\cal W} (z)  \approx   
8 \sum_{n=1}^\infty \frac{1-\cos (2 nz  )}{n
 }  = 8 \ln \left| \frac{\sin\left(z + \frac{iT }{D_0} \right)}{\sinh \left(T/D_0\right) }\right|,
 \label{appe.3.6.bis}
 \eneq
 with the cutoff $D_0$ reintroduced at the last step in Eq.~(\ref{appe.3.6.bis}) to 
 assure the convergence of the sum over $n$ for all $z$.
As final step, which we also employ in computing $G(\omega)$, 
we recall the periodicity (with period $\beta$) of the integrand 
of the above integrals over $\tau$. 
We thereby arrive at a compact expression for the partition function ${\cal Z}$ in the large-$\Gamma$
limit,
\begin{widetext}
\begin{equation}
{\cal Z}  = {\cal N}   \sum_{m=-\infty}^\infty \: e^{-\frac{\beta v}{2 \pi L} \left[\alpha 
 +2 \pi \left(m + \frac{1}{2} -S_z s_\sigma \right)\right]^2}
 +  \frac{Y^2\beta^2 D_0^2 {\cal N}  }{ \pi}
 \int_{-\infty}^\infty  d z \: 
 e^{-\frac{2 v \beta}{\pi L}\left[z + \frac{ \pi + \alpha -2 \pi S_z s_\sigma }{2} \right]^2 -{\cal W} \left(z \right)} . 
 \label{appe.3.7}
 \end{equation}   
\end{widetext}

Following the same path leading to Eq.~(\ref{appe.3.7}), we next compute the 
  instanton corrections to the linear AC conductance. In order to do so, in Eq.~(\ref{appe.3.1}), we decompose
  $\tilde{w}_- (i\Omega)$ into a saddle-point solution $\tilde{w}_{\rm sp}(i\Omega)$ plus
  fluctuations $\delta \tilde{w}_- (i\Omega)$. Here, $\tilde{w}_{{\rm sp},-}(i\Omega)$ either corresponds to a uniform solution, $\tilde{w}_{{\rm sp},-} (i\Omega)=\beta w_{-,m} \delta_{\Omega,0}$, or to   
  a single instanton--anti-instanton (``$+$'' sign) or anti-instanton--instanton (``$-$'') pair, where Eqs.~\eqref{appe.3.3} and \eqref{appe.3.4} give
\beq\label{fluc}
  \tilde{w}_{-}^{(m)}(i\Omega) = \beta w_{-,m} \delta_{\Omega,0} \pm \frac{2\pi}{i\Omega} \left( e^{-i\Omega \tau_1} - e^{-i\Omega \tau_2} \right).
\eneq 
 Expanding up to second order in  $\delta \tilde{w}_- (i\Omega)$, the fluctuation action is given by
\beq
  {\cal S} [ \delta \tilde w_- ] = \frac{T}{2} \sum_{i\Omega} \left[ {\cal K}_{-,-}(i\Omega)+\Gamma \right]
  \delta \tilde{w}_- (-i\Omega)  \delta \tilde{w}_- (i\Omega) . 
  \label{appe.3.8}
\eneq
As ${\cal S}[\delta\tilde w_- ]$ is quadratic in the fluctuations, we can integrate over the field
$\delta\tilde w_-$. The off-diagonal elements of the kernel
${\cal K}_{a,a'} (i\Omega)$ in Eq.~\eqref{appe.1.4}, with a corresponding structure of the inverse
kernel, imply two remarkable effects discussed next.

First, ${\cal K}_{a,a'}  (i\Omega)$ is modified to 
\beq\label{kgamma}
[{\cal K}_\Gamma]_{a,a'} (i\Omega)={\cal K}_{a,a'}(i\Omega) -
\frac{ {\cal K}_{a,-} (i\Omega) {\cal K}_{-,a'} (i\Omega)}{
{\cal K}_{-,-} (i\Omega) +\Gamma},
\eneq
see Eq.~\eqref{appe.3.8}.  As a consequence of this effect, 
to leading order in the frequency $\omega$, 
the linear AC conductance $G_* (\omega)$ is given by
\beq
G_*(\omega) = \frac{e^2}{2\pi}  + i L_{\rm kin}^\Gamma \omega + {\cal O} (\omega^2) , 
\label{appe.3.9}
\eneq
with the renormalized kinetic inductance 
\beq \label{lkingam}
L_{\rm kin}^\Gamma = \frac{e^2}{\pi v} 
\left(L_c+W-L+\frac{L}{1+\pi \Gamma L} \right).
\eneq
We note in passing that for $\Gamma =0$, even though our derivation is not valid in that case,
Eq.~\eqref{lkingam} predicts $L_{\rm kin}^{\Gamma=0}=L_{\rm kin}$, where $L_{\rm kin}$ in Eq.~(M9) 
pertains to the small-$\Gamma$ case.  We next show that one also obtains an
effective capacitance contribution due to EVT in the large-$\Gamma$ case.

Second, through the effective action ${\cal S}[\tilde w_+,\tilde w_-]$ in Eq.~\eqref{sw+w-}, 
the field $\tilde w_+(i\Omega)$, which ultimately determines the conductance according to Eq.~(M8), couples
to the saddle-point solutions $\tilde w_{-}^{(m)}(i\Omega)$ in Eq.~\eqref{fluc}.  
Using the kernel $K_\Gamma$ in Eq.~\eqref{kgamma}, we obtain
\begin{eqnarray}
\langle \tilde{w}_+ (-i\Omega)\tilde{w}_+ (i\Omega) \rangle &=& \frac{1}{[{\cal K}_\Gamma]_{+,+}(i\Omega)} 
\label{appe.3.10} \\
&+&  T \left|\frac{[{\cal K}_\Gamma]_{+,-} (i\Omega)}{[{\cal K}_\Gamma]_{+,+}(i\Omega)}\right|^2
\left|\tilde{w}_{-}^{(m)}(i\Omega)\right|^2 \nonumber. 
\end{eqnarray}
Using Eq.~\eqref{fluc}, we eventually arrive at
\begin{widetext}
\beq
\langle \tilde{w}_+ (-i\Omega)\tilde{w}_+ (i\Omega) \rangle = \frac{1}{[{\cal K}_\Gamma]_{+,+}(i\Omega)} 
+ 
 \frac{4 \pi \beta D_0^2 Y^2}{   {\cal Z}  }   \int_{-\infty}^\infty  d z 
\frac{ 1-\cos \left(\frac{\beta \Omega z}{\pi}\right) }{\Omega^2}  \,
\left| \frac{[{\cal K}_\Gamma]_{+,-} (i\Omega) }{[{\cal K}_\Gamma]_{+,+} (i\Omega)}\right|^2 
 e^{-\frac{2  v \beta}{ \pi L} \left[z + \frac{\pi + 
 \alpha  -2\pi S_z s_\sigma}{2}\right]^2  - {\cal W} \left(z\right)} .
\label{appe.3.11}
\eneq
\end{widetext}
\noindent
We are now ready to perform the analytic continuation of ${\cal G}(i\Omega)$ in Eq.~(M8) to real 
frequency $\omega$.  
By expanding to lowest order in $\omega$,  Eq.~(\ref{appe.3.11}) thereby gives the 
AC conductance in the large-$\Gamma$ limit (with $(-1)^{n_v}\to 1$) as
\beq
G_*(\omega) =\frac{e^2}{2\pi} + i (L_{\rm kin}^\Gamma - C_{\rm eff}^\Gamma) \omega + {\cal O}(\omega^2)  , 
\label{appe.3.12}
\eneq
with the kinetic inductance $L_{\rm kin}^\Gamma$ in Eq.~(\ref{appe.3.9}) and the effective capacitance
\beq
C_{\rm eff}^\Gamma =   \frac{e^2 \beta^3 D_0^2 Y^2}{ \pi^3  {\cal Z}  } 
\int_{-\infty}^\infty  d z \, z^2  e^{-\frac{2  v \beta}{ \pi L} \left[z + \frac{\pi + 
\alpha -2 \pi S_z s_\sigma }{2}\right]^2  - {\cal W} \left(z\right)
 }    .
 \label{appe.3.13}
 \eneq
Using this expression, we arrive at Eq.~(M10). 
 
 An important observation that arises from comparing the expressions for $G(\omega)$ for
 small vs large $\Gamma$ is that $G(\omega=0)=\frac{e^2}{2\pi}$ is identical in both regimes. 
 In Sec.~III, we provide a general argument implying that a finite $\Gamma$ never changes 
 $G(0)$.
 Therefore, the EVT rate $\Gamma$ can only change frequency-dependent contributions to the conductance in our
 setup, regardless of the  value of $\Gamma$. 
 
 \section{III. On the DC current }
\label{absence}

We reported above (and in the main text) that $G(0)$ is not changed
by a finite EVT rate $\Gamma$ in our setup, neither for small $\Gamma$,
see Sec.~II B, nor for large $\Gamma$, see Sec.~II C.
In what follows, we prove the absence of zero-frequency corrections 
to the total current $I$ due to $\Gamma$ under a constant applied voltage bias $V$. 
As discussed in the main text, the physical reason for this result is the chiral
anomaly of the chiral Majorana edge states.

Rather than resorting to linear response theory,
let us consider the imaginary-time action for the fields $w_\pm(\tau)$ in the presence of 
a finite voltage $V(\tau)$, denoted by ${\cal S}_{\rm v} [w_+,w_-]$. Setting for simplicity the fermion tunneling amplitudes $\lambda_1=\lambda_2=0$, we find
\begin{eqnarray}
&& {\cal S}_{\rm v} [w_+,w_-] = \frac{T}{2} \sum_{i\Omega} \: \sum_{a,a'}
 {\cal K}_{a,a'}  (i\Omega)\tilde{w}_a (-i\Omega) \tilde{w}_{a'} (i\Omega)\nonumber \\
 &&-\frac{eT}{2\pi} \: \sum_{i\Omega}\tilde{V}(-i\Omega) \tilde{w}_+ (i\Omega) + S_\Gamma [w_-] ,
\label{appe.4.1}
\end{eqnarray}
where $\tilde{V} (i\Omega)$ is the Fourier-Matsubara transform of $V(\tau)$.
We recall that $S_\Gamma$ depends only on $w_-$. 
Using Eq.~(M4), we  obtain the Fourier-Matsubara components of the current as 
\beq
\tilde I(i\Omega) = - \frac{ie}{\pi} \int_0^\beta d \tau  e^{-i\Omega \tau}
\left\langle \frac{d \tilde{w}_+ (\tau)}{d \tau}\right\rangle
= \frac{\Omega e}{\pi} \langle \tilde{w}_+(i\Omega) \rangle . 
\label{appe.4.2}
\eneq
For arbitrary $\Gamma$, the right-hand side of Eq.~(\ref{appe.4.1}) is quadratic in $w_+$. 
This fact allows us to functionally integrate over $w_+$, resulting in an effective action 
$\tilde{\cal S}_{\rm v}[w_-]$ which only depends on $w_-$.  Explicitly, we find
 \begin{eqnarray}
 \tilde{\cal S}_{\rm v} [w_-] &=& \frac{T}{2} \sum_{i\Omega} {\cal K}_{-,-} (i\Omega)\tilde{w}_- (-i\Omega)
 \tilde{w}_- (i\Omega) +  S_\Gamma \nonumber \\
 &+&\frac{T}{2}\sum_{i\Omega} \frac{ \left|{\cal K}_{+,-}(i\Omega)\tilde{w}_- ( i\Omega)-\frac{e\tilde{V} ( i\Omega)}{2\pi}\right|^2
  }{{\cal K}_{+,+}(i\Omega)}. 
   \label{appe.4.3}
\end{eqnarray}
Differentiating the corresponding generating functional with respect to $\tilde{V} (i\Omega)$,  we 
obtain a formally exact expression for the current,
\beq
\tilde I(i\Omega) = \frac{\Omega e^2}{2\pi^2} \left(\frac{\tilde{V} (i\Omega)}{{\cal K}_{+,+}(i\Omega) }-
\frac{{\cal K}_{+,-} (i\Omega) \langle \tilde{w}_- (i\Omega)\rangle }{{\cal K}_{+,+} (i\Omega) }\right). 
\label{appe.4.4}
\eneq
From Eq.~(\ref{appe.1.4}), for $\Omega \to 0$, we find by direct inspection
\beq
\frac{{\cal K}_{+,-} (i\Omega)}{{\cal K}_{+,+} (i\Omega) } \sim \Omega L, \label{appe.4.5}
\eneq
as well as 
\beq
[{\cal K}_{+,+}(i\Omega)]^{-1}  = \frac{\pi}{|\Omega|} + {\cal O} (|\Omega|^0).
\eneq
Inserting these relations into Eq.~(\ref{appe.4.4}) and taking the limit
$\Omega \to 0$, we observe that only the first term on the right-hand side of Eq.~(\ref{appe.4.4}) 
contributes to the DC current.  However, this term is 
independent of $\tilde{w}_-$ and, therefore, is blind to the EVT rate $\Gamma$. We arrive at
the general conclusion that a finite EVT rate $\Gamma$ can neither affect the DC current nor the DC conductance $G(0)$ in our setup at any order in the applied voltage $V$.

\bibliography{biblio}

\end{document}